# wtest: an R Package for Testing Main and Interaction Effect in Genotype Data with Binary Traits

Rui Sun[1,2], Billy Chang[1], Benny Chung-Ying Zee[1,2,*] and Maggie Haitian Wang[1,2,*]

[1]Division of Biostatistics and Centre for Clinical Research and Biostatistics (CCRB), JC School of Public Health and Primary Care, the Chinese University of Hong Kong, Hong Kong SAR

[2]Centre for Clinical Trials and Biostatistics, CUHK Shenzhen Research Institute, Shenzhen, China

* Correspondence: maggiew@cuhk.edu.hk

**Abstract**

This R package evaluates main and pair-wise interaction effect of single nucleotide polymorphisms (SNPs) via the W-test, scalable to whole genome-wide data sets. The package provides fast and accurate p-value estimation of genetic markers, as well as diagnostic checking on the probability distributions. It allows flexible stage-wise or exhaustive association testing in a user-friendly interface.

**Availability:** The package is available in CRAN, or from website: http://www2.ccrb.cuhk.edu.hk/wtest

**Background**

Single nucleotide polymorphism (SNPs) association testing is an important step for selecting potential drug target and understanding gene function in disease etiology. Current methods mainly incorporate classical probability distributions or permutation tests to calculate p-values. However, tests based on theoretical probability distributions might have affected power in the challenging genetic environment of sparse data and complicated genetic architectures; on the other hand, tests using permutations face heavy computing burden to reach genome-wide significant threshold of 5.1 x 10$^{-8}$ and the requirement is becoming more stringent with the increasing sequencing depth (Prokopenko et al., 2009; Sebastiani et al., 2012). In this paper, we introduce a software that provides fast and bias-corrected p-value calculations for main or pairwise interaction effect in genotype data sets. The package also provides user-friendly options to perform exhaustive or stage-wise interactions evaluations and diagnostic checking. It offers an integrative and practical tool to evaluate genetic association in large genetic data sets.

**Method overview and implementation**

The wtest package is based on the W-test (Wang, Sun et al.), which measures the association between binary phenotype and categorical genotype data. It tests for the null hypothesis that there is no distributional difference of a subset in the case group and in the control group. The test statistic takes the following form,

$$W = h \sum_{i=1}^{k} \left[ \log \frac{\hat{p}_{1i}/(1-\hat{p}_{1i})}{\hat{p}_{0i}/(1-\hat{p}_{0i})} \Big/ SE_i \right]^2 \sim \chi_f^2$$

where $n_{1i}$ and $n_{0i}$ are the number of cases and controls in the i$^{th}$ cell; $p_{1i} = n_{1i}/N_1$ and $p_{0i} = n_{0i}/N_0$ are the conditional cell probabilities of the i$^{th}$ cell; k denotes the number of non-empty categories formed by a subset; and $SE_i$ is the standard error of the i$^{th}$ log odds ratio. The W-test follows a chi-squared distribution of f degrees of freedom. The scalar h and degree of freedom f take forms of covariance matrices of the log odds ratios, which are estimated from bootstrapped samples under the null hypothesis by large sample theories. Therefore, the W-test inherits a data-set adaptive degree of freedom that absorbs the genetic variation not attribute to phenotypes, and the test is therefore robust in complicated genetic architectures. A more detailed description can be found in Wang, Sun et al. (2016).

The implementation of W-test contains two main steps: 1) Estimation of h and f; and 2) W-test p-value calculation. The usage of the package is illustrated by an example dataset included in the package, which contains 115 diabetic individuals and 23 SNPs (Wang et al., 2014).

Step 1. Estimation of h and f by hf.calculation: this function estimates the h and f values using bootstrapped samples and permutated phenotypes. If skipped, the Step 2 W-test will be calculated using default hf values. When sample size is large and genetic architecture is uniform, the estimated and default values will get close. The hf.calculation function automatically produces a matrix of h and f for each k. The estimated h and f of the example data can be found in Supplementary Materials (Table S1 and S2). Simulation studies using 1,000 variables and 1,000 subjects showed that the h and f estimates will converge when B is greater than 200 (Figure S1). The user has the option to increase the number of bootstrap times B, recommended when the data dimension is smaller.

Step 2. W-calculation by wtest: This function outputs the W-test statistics and p-values with a given hf matrix. The user needs to indicate the desired the order of association to evaluate: order = 1 for main effect, and order = 2 for interaction effect. An option is provided to control the number of input variables by ordering their p-values through the input.poolsize or input.pval arguments. Similarly, the number of output markers can be filtered by the output.pval argument. For example, by setting order = 2, input.pval = 0.5 and output.pval = 0.003, the function will calculate epistasis among the SNPs that main effect p-values are smaller than 0.5, and will output the pairs with p-value < 0.003. The function has the following output:

**Table 1.** Interaction effect output from example data

|   | SNP1 | SNP2 | W | k | Pair-pval | SNP1 pval | SNP2 pval |
|---|------|------|---|---|-----------|-----------|-----------|
| 1 | rs2383207 | rs3772622 | 31.1 | 9 | 7.3E-4 | 0.123 | 0.002 |
| 2 | rs1014290 | rs3772622 | 28.5 | 9 | 1.9E-3 | 0.131 | 0.002 |
| 3 | rs10012946 | rs3772622 | 18.9 | 6 | 2.7E-3 | 0.295 | 0.002 |

Pair-pval, SNP1 and SNP2 pval: p-value of the pair, SNP1 and SNP2, respectively.

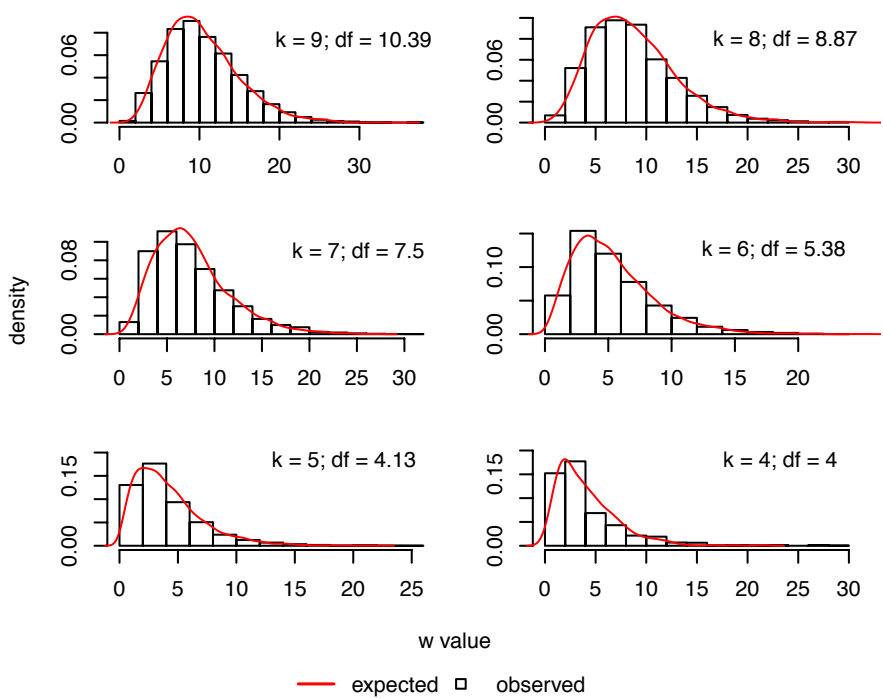

**Fig. 1.** Diagnosis density plot

Diagnosis test can help the user to examine the appropriateness of the testing probability distributions, a feature that is important in genetic analysis. In this package, we provide two diagnoses in visualizations: the density plots and the QQ-plot. The function w.diagnosis plots the histogram of W-values obtained by real genotypes under the null hypothesis (observed), and the density curve of chi-squared distribution of f degrees of freedom (expected) (Figure 1). Close overlaying of the two densities infers that the estimated h and f give appropriate probabilities for conducting hypothesis testing. A panel representation indexed by k provides detailed check on the distributions at each degree of freedom.

Computing efficiency: The wtest R package incorporates C codes to increase computing efficiency. The speed is evaluated on a laptop computer of 1.6GHz Intel Core i5 processor and 4GB RAM. On a data set consists of 5,000 subjects and 100 SNPs, when B= 200, n.sample = 1000, the time elapsed for estimating h and f is 40.5 s. With a given h and f matrix, the time used to compute main effect is 0.042s. For exhaustive interaction evaluation, wtest used 1.69s, while the chi-squared test used 36.41 and logistic regression used 130.56, on the same data set . The wtest is around 20-fold faster than the chi-squared test and 76-fold faster than the logistic regression in R. For the calculation of genome-wide main effects on a laptop computer, wtest took 12.67s on one chromosome containing 20,000 SNPs and 5,000 subjects. Accordingly, for genome-wide calculation reaching 500,000 SNPs, the estimated time is around 5 minutes. C++ software is available for faster exhaustive interaction effect calculation in whole genome-wide data [ref: http://www2.ccrb.cuhk.edu.hk/wtest/download.html]. The R package offers a convenient and flexible analysis tool for screening and selecting SNPs in large data set in the R environment.

The wtest package integrates main effect and pairwise interaction evaluations in genotype data with binary traits, and allows explorations of the data in flexible ways including exhaustive, stage-wise or on selected variables. The p-value calculation is fast and free from time-consuming permutations; and the data-adaptive probability distributions give p-values corrected from bias due to sparse data and complicated genetic architectures. The package also includes handy diagnosis checking tools. As a whole, the software is a powerful and convenient tool for both non-experts and professionals to analyze disease associations in genotype data of small or large scale.

**Acknowledgements**

This research was conducted using the resources of the High Performance Cluster Computing Centre, Hong Kong Baptist University, which receives funding from Research Grant Council, University Grant Committee of the HKSAR and Hong Kong Baptist University.

**Funding**

This study is supported by NSFC [81473035, 31401124] to MH Wang.

**Competing interest**

The authors declare no competing interest.